# An omniscient Maxwell's demon

Aishwarya Kumar, Tsung-Yao Wu, Felipe Giraldo Mejia, David S. Weiss[*]

Department of Physics, The Pennsylvania State University, University Park, PA 16802

In 1872, Maxwell proposed his famous demon gedanken experiment[1]. By discerning which particles in a gas are hot and which are cold, and then performing a series of reversible actions, Maxwell's demon could rearrange the particles into a manifestly lower entropy state. The apparent violation of the second law of thermodynamics was resolved in the twentieth century[2]: Maxwell's demon often increases the entropy of the universe while gathering his information[3], and there is an unavoidable entropy increase associated with the demon's memory[4]. Despite its theoretical resolution, the appeal of the demon construct has led many experiments to be framed as demon-like. However, past experiments have either had no intermediate information storage[5], negligible change in the system entropy[6,7], or have involved systems of four or fewer particles[8-10]. Here, we present an experiment that realizes the full essence of Maxwell's demon. We start with a randomly half-filled 3D optical lattice with ~60 atoms. We can make the atoms sufficiently vibrationally cold that the initial disorder is the dominant entropy. After determining where the atoms are, we execute a series of reversible operations to create a fully filled sub-lattice, a manifestly low entropy state. Our Maxwell demon lowers the total entropy by a factor of 2.44, enough to cool the ensemble below the quantum degeneracy threshold[11]. We plan to use this Maxwell demon to initialize a neutral atom quantum computer.

---

[*] Corresponding Author. dsweiss@phys.psu.edu

There has been a recent boom in cold atom sorting experiments. Atoms in a variety of arrays of dipole light traps have been impressively rearranged by moving individual traps [6,7,12,13]. The entropy associated with disordered occupancy in those cases is at most about 10% of the system entropy[6], which is dominated by vibrational excitation in the traps. Four atoms in a 1D optical lattice [8] have been compacted using a method[14] similar to the one we demonstrate here with 50 atoms in 3D.

Our experiment proceeds as follows. We prepare a randomly 56% filled blue-detuned 3D lattice with 4.9 μm lattice spacing [15]. By imaging polarization gradient cooling laser light, we determine occupancy across the lattice with $<10^{-6}$ error/site in 800 ms[15]. Projection sideband cooling [16] puts 89% of the Cs atoms into their vibrational ground states and >99.7% of them in the $|F=4, m_F=-4\rangle$ hyperfine ground state. We then combine the ability to address atoms at individual sites, using crossed laser beams and microwaves to make site-dependent state changes[17], with the ability to make state-dependent lattice translations, by rotating lattice beam polarizations[18]. Starting from a given 3D occupancy map we devise a sequence of operations to fill up either a 5×5×2 or a 4×4×3 sub-lattice.

We can target any site in a 5x5x5 lattice using a pair of focused addressing beams intersecting at a right angle[17,19]. Targeting proceeds as in our previous demonstration of high fidelity single qubit gates, but the magnetic sublevels are different and in this case we are unconcerned with long term quantum coherence. The addressing beams shift the ($|F=4, m_F=-4\rangle$ to $|F=3, m_F=-3\rangle$ resonance by ~50 kHz, which allows us to drive the associated microwave transition using an adiabatic fast passage pulse (AFP, see Methods for details) that transfers only the target atom. An atom making the transition from $m_F=-4$ to $m_F=-3$ moves from the "stationary" to the "motion" state.

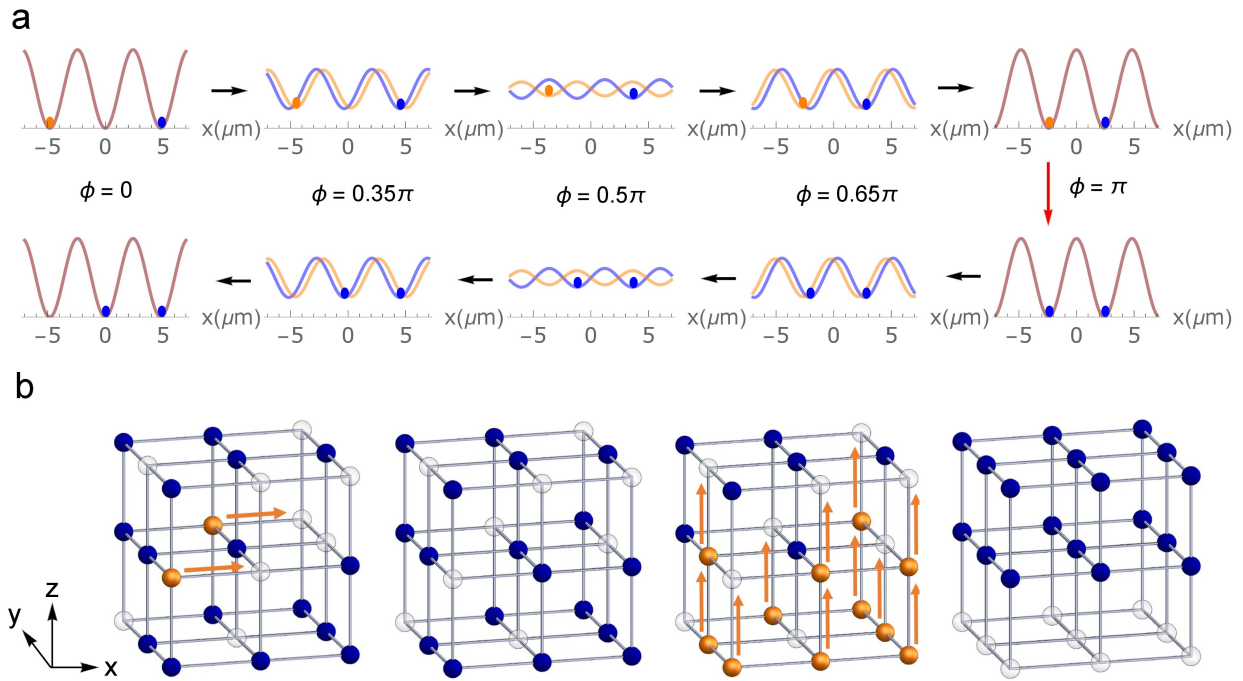

**Figure 1. Motion step and sorting algorithm. a,** Motion steps to fill a vacancy in a given direction. The curves show the lattice potential as a function of position for the "motion" state (orange curve) and the "stationary" state (blue curve). The arrows denote a time series as the angle ($\phi$) between the polarizations of the two lattice beams is adiabatically ramped to $\pi$ and back. The atom to be moved is transferred to the "motion" state (orange circle) using targeted addressing at the beginning. As the polarization of one of the lattice beams is rotated, the atoms in the motion state and the "stationary" state (blue circle) move in opposite directions, settling half a lattice space away from their original positions when $\phi = \pi$. The atom in the motion state is then optically pumped to the stationary state (illustrated by the red arrow). As the polarization is rotated back, both the atoms move in the same direction, with the atom that started in the stationary state returning to its original position and the atom that started in the motion state moving by one lattice site. **b,** A simplified illustration of two parts of the sorting algorithm in a 3x3x3 lattice. The first motion step "balances" the array such that every z-row has exactly 2 atoms. The second motion step "compacts" atoms into two planes.

The linear polarizations of the two beams that create the lattice in a given direction are initially aligned, so the two states are trapped nearly identically. When the polarization of one of the lattice beams is rotated, using two electro-optic modulators and a λ/4 plate, the optical lattices for the two states move in opposite directions (see Figure 1a). After we rotate by π, we optically pump the atoms back to the stationary state and rotate the polarization back. The net effect of this sequence is that atoms that start in the stationary state move but return to the same place, while atoms that start in the motion state are shifted by one lattice site.

The sorting algorithm for compacting atoms in the lattice was proposed in previous work[14,20]; we have slightly modified it to allow the filling of any continuous sub-lattice (see Methods). The general idea is to first perform a series of balancing steps in the $x$ and $y$ directions so that every row in the $z$ direction has the required number of atoms to fill a desired number of planes. Then a series of compaction steps in the $z$ direction moves atoms to fill the planes of the target sub-lattice (Figure 1b). For example, for filling a 5x5x2 sub-lattice from a half filled 5x5x5 lattice, atoms are first "balanced" in the $x$ and $y$ directions such that every row in the $z$ direction has at least 2 atoms. Parallel $z$ motion steps then move the atoms to the desired planes. After sorting, we reimage the atoms, and repeat the procedure to correct any errors. The ability to know exactly where the vacancies are is an advantage of this approach to filling a lattice compared to implementing a superfluid-Mott insulator transition[21], where residual occupancy errors are unknown.

Figure 2 shows two implementations of this algorithm in which target sub-lattices were completely filled after two sorts. In general, starting with at least half the lattice sites filled in a 5x5x5 array, three sorts leave us with an average filling fraction of 0.97 (0.95) and perfect filling rate of 0.32 (0.27) for 5x5x2 (4x4x3) target sub-lattices. For the first sort, the average number of motion steps was 6.4 (5.6) and the average number of addressing operations was 38 (62) for filling a 5x5x2 (4x4x3) sub-lattice. Each sort takes ~190 ms on average. Figure 3 shows the filling fraction as a function of the number of sorts. These numbers match well with Monte Carlo simulations that consider measured sources of error (see Methods). A major source of error for atoms in both motion and stationary states is spontaneous emission from the lattice. The spontaneous emission rate is significantly higher (17 times on average) during a motion step because the lattice intensity is not zero at the trap minima during the motion (see Figure 1a). When an atom spontaneously emits a photon and changes hyperfine state, it becomes anti-trapped and is lost. The measured average loss per motion step is $\sim 4\times 10^{-3}$. Another source of error is imperfect transfer of atoms from the stationary state to the motion state. Our measured transfer fidelity is 0.986, limited by a combination of imperfect addressing beam shape, pointing noise of the addressing beams and magnetic field fluctuations. This error can cause two atoms to end up in the same lattice site, both of which are lost during imaging. The number of sorts that can be performed to fill errors is eventually limited by the 92 s vacuum lifetime and double atom loss. Optical pumping leads to a modest amount of heating, exciting about 7% of the population from the 3D vibrational ground state per motion step. Were we to replace the more convenient optical pumping with targeted addressing, this number would be reduced to 0.6%.

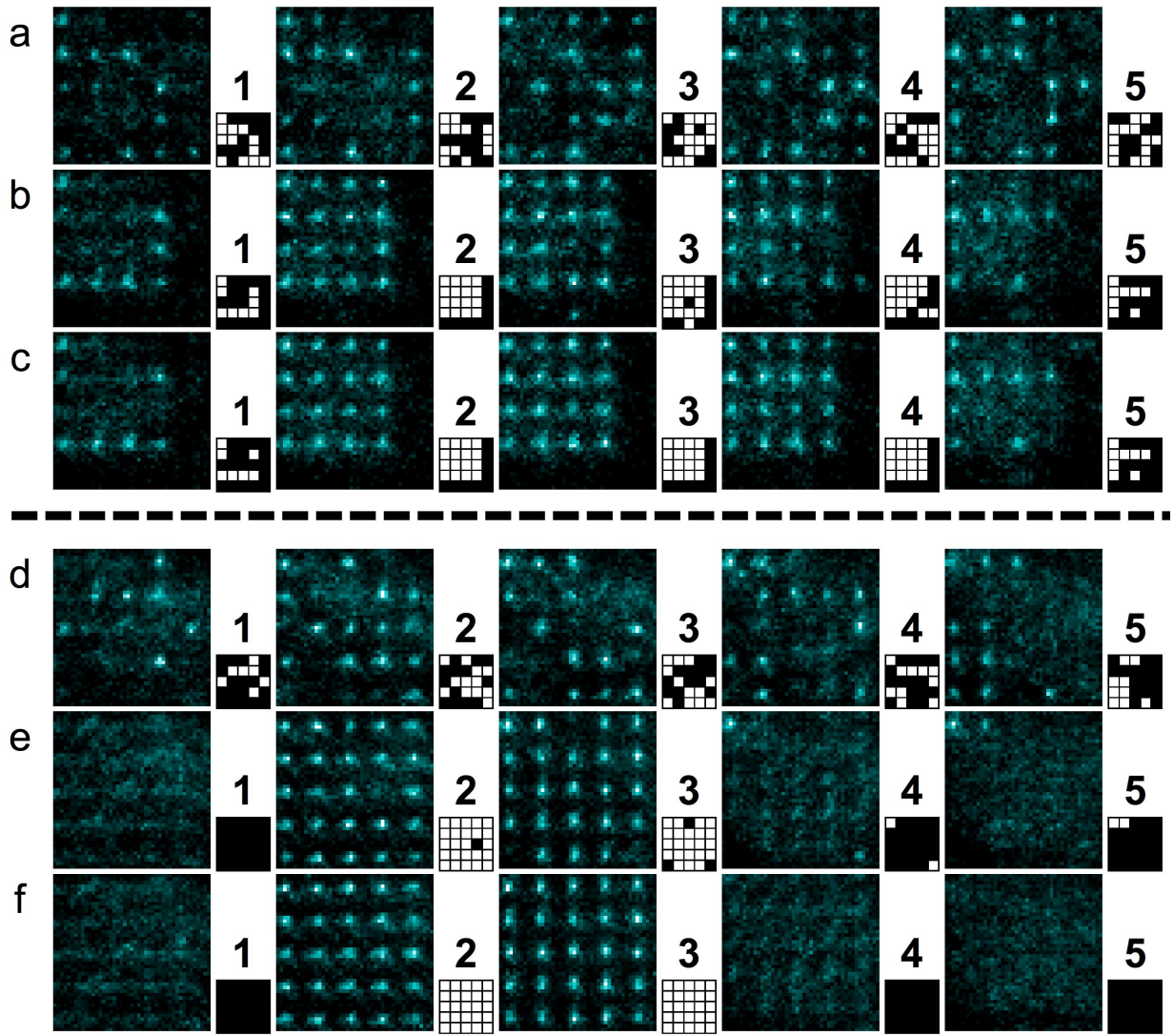

Figure 2. Perfect filling of 4x4x3 and 5x5x2 sub-lattices. a, b, c, Two sorts to fill arrays of 4x4x3 and d, e, f, 5x5x2. Images are taken at the beginning of the sequence (a, d) after occupancy maps are created. The 5 images in each row show different planes of the lattice in focus. The insets are the real-time computer-generated occupancy maps. The motion steps are then calculated by the sorting algorithm and executed in real-time. Then another set of pictures (b, e) is taken. Finally, a second sorting algorithm is applied to fill any errors, and a final occupancy map is generated (c, f). Atoms outside of the target sub-lattice can be kept as spares, as in c1 and c5, or they can be selectively state-flipped and removed by a resonant clearing beam. The absence of spare atoms in f is coincidental.

After sorting, and a final round of projection cooling, we measure the vibrational sidebands to determine the final ground state occupation, as shown in Figure 4. Projection sideband cooling (see Methods) leads to ground state occupation probabilities of 0.949(7), 0.954(6) and 0.985(1) in the *x*, *y* and *z* directions respectively, which implies 89% occupation of the 3D vibrational ground state. The state is not quite thermal, but the population is overwhelmingly in the lowest three levels. We calculate that the vibrational entropy for this state is ~0.59$k_B$ per particle.

The configurational entropy is given by[11]:

$$S = \frac{1}{\bar{n}}\left(\bar{n}\ln\left(\frac{1}{\bar{n}}\right) + (1-\bar{n})\ln\left(\frac{1}{1-\bar{n}}\right)\right), \tag{1}$$

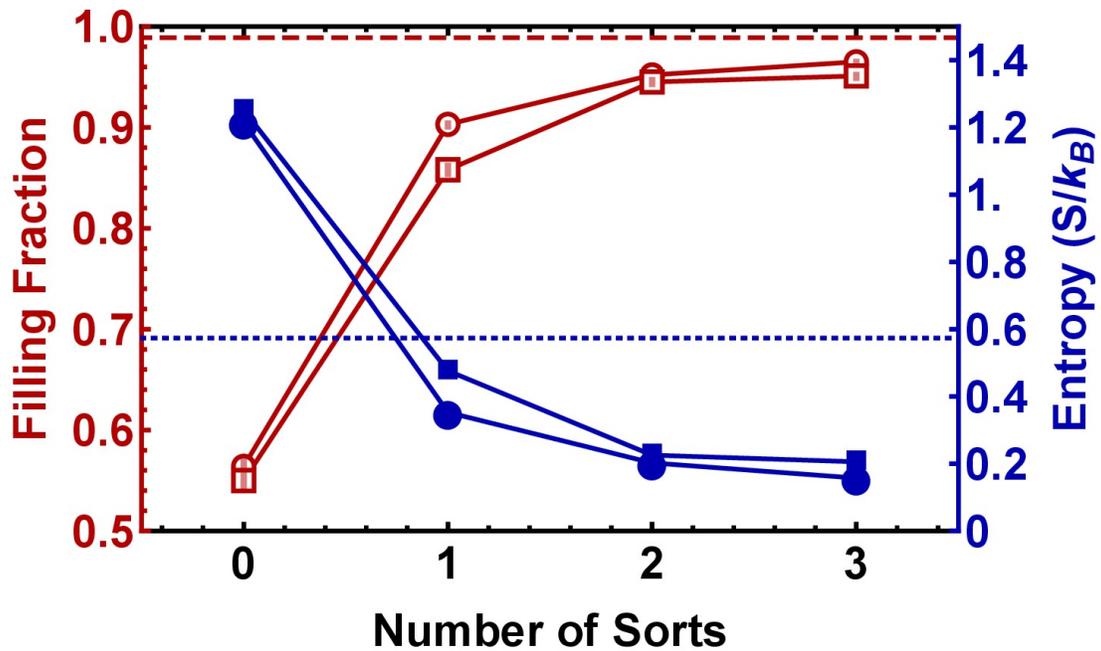

Figure 3. Filling fraction and entropy. The red hollow markers show the filling fraction as a function of the number of sorts for 5x5x2 (circles) and 4x4x3 (squares) target sub-lattices. The red horizontal dashed line is the limit associated with loss from collisions with background gas atoms during the 1 s time it takes to image and sort. The blue solid markers show the configurational entropy/atom as a function of the number of sorts for 5x5x2 (circles) and 4x4x3 (squares) target sub-lattices. The total entropy at the beginning and at the end is the sum of the vibrational entropy (blue horizontal dotted line) and the configurational entropy; sorting reduces it by a factor of 2.44.

where $\bar{n}$ is the filling fraction. The solid blue line in Figure 3 shows the configurational entropy as a function of the number of sorts, and the dotted line shows the vibrational entropy after projection cooling. Sorting reduces the configurational entropy by a factor of 8 and the total entropy by a factor of 2.44. The final total entropy per particle, $0.75k_B$, is comfortably below the entropy per particle ($1.24k_B$) required for the system to pass the Bose Einstein condensation threshold when the lattice is adiabatically shut off and the atoms are left in a 3D box potential[11].

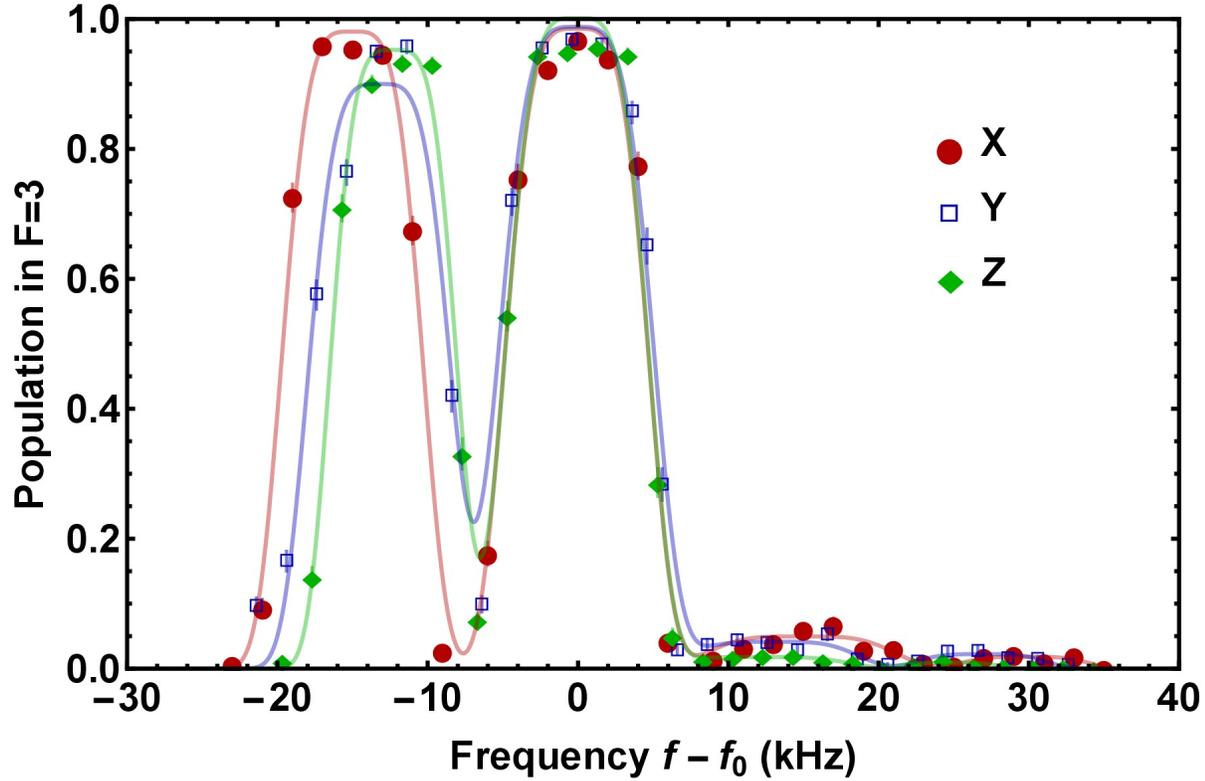

**Figure 4.** Microwave spectra after projection cooling. After projection cooling, the atoms are in the $|F = 4, m_F = -4\rangle$ state. A projection between different vibrational states is then induced by rotating the polarization of a lattice beam in one of the x, y or z directions, and an AFP microwave pulse is applied to transfer the atoms to $|F = 3, m_F = -3\rangle$. A resonant clearing beam then pushes all the atoms in F = 4 out of the lattice. The remaining population is plotted as a function of the difference between the microwave frequency (f) and the carrier frequency ($f_0$). The solid maroon circles give the spectrum for the x direction, hollow blue squares for the y and solid green diamonds for z. The maroon, blue and green solid lines are sums of 4 fitted super-Gaussians of order 4 for the x, y, and z directions respectively. The figure shows that the 3D vibrational ground state, which does not contribute to the rightmost sidebands, is occupied 88.9(9)% of the time.

The number of required motion steps scales as $N^{1/3}$ where $N$ is the number of atoms to be sorted. Monte Carlo simulations, assuming our current error rate, show that starting from a half filled 10x10x10 lattice, 10x10x4 and 7x7x7 sub-lattices can be filled to a ~ 0.93 filling fraction. The error due to the motion could be reduced by further detuning the lattice light. Tripling the detuning would decrease the spontaneous emission rate by a factor of 9 and the lattice depth by a factor of 3. Moving atoms three times slower, the total spontaneous emission would be reduced three-fold, which would improve the filling fraction to ~0.975 for ~400 sorted atoms. It should be possible to improve microwave transfer errors by an order of magnitude by improving magnetic field stabilization and adapting our phase gate[19], which is insensitive to addressing beam intensity fluctuations.

Since atoms in our 3D lattice geometry have many more near neighbors than in other geometries (including 3D arrays of dipole traps[12], where the traps need to be far apart in the third dimension), small known filling errors can be more readily incorporated into the design of any quantum computation. With coherence times > 15 s and the possibility of much better cooling deeper in the Lamb-Dicke limit, this experiment improves the prospects for creating large scale entanglement through cold-collision gates[22,23] and implementing one way quantum computation[24]. It might also allow for exploration of Ising spin Hamiltonians in three dimensions using Rydberg atoms[25,26].

Laser cooling processes generally involve selectively acting on particles differently depending on their momentum, energy or internal state[5,27-31]. Because there is no stored information, these mechanisms differ in spirit from Maxwell's demon, and other gedanken demons[32]. The entropy increase of the outside world is built into the cooling cycle, visible in lost particles or scattered light. These experiments do not evoke the theoretical paradox that 20$^{th}$ century information theory worked to resolve[2].

Our experiment, in contrast, is conceptually like the homunculus Maxwell envisioned. Collected information is stored and used as a guide to the execution of reversible operations. One difference in our experiment is that Maxwell's demon collected information one particle at a time, acting on each in turn. Our demon is omniscient after obtaining an occupancy map of the whole system, so that it can map out a plan to act on the particles in parallel to put them in a manifestly low entropy state. Maxwell's gedanken demon led to a deep understanding of the relationship between entropy and information. Our practical omniscient Maxwell demon has demonstrated a new way to pass the quantum degeneracy threshold in an atomic gas and to initialize a neutral atom quantum computer.

**Acknowledgements** This work was supported by the U.S. National Science Foundation grant PHY-1520976.

**Author Contributions** All the authors contributed equally to this work.

**Competing Financial Interests** The authors declare no competing financial interests.

**Materials and Correspondence** The data presented in this report are available on request to D.S.W.

# Methods

## Apparatus

We load atoms from a Magneto-Optical Trap to a 3D optical lattice formed by 3 pairs of 75 μm waist, 838.95 nm laser beams. Each lattice beam has 250 mW, giving a lattice depth of 190 μK at the central lattice site. The two beams in each pair cross each other at $10^o$, yielding a 4.9 μm lattice spacing. Two pairs are frequency shifted relative to the third by +30 and -175 MHz, to prevent mutual interference among the lattice pairs. One beam in each pair has two Pockels cells in its path aligned so that their axes are at 45 degrees relative to the incoming polarization, followed by a λ/4 wave-plate aligned with the incoming polarization. As the voltage on the Pockels cells is increased, the polarization of this beam rotates. The angle of rotation is π when the half-wave voltage is applied to both Pockels cells.

## Projection cooling improvement

We have improved the performance of our previously demonstrated projection sideband cooling [16], from 76% to 89% ground vibrational state occupancy. The improvement results from two changes. First, we increased the fidelity of the Δv = -2 microwave pulse, where Δv is the microwave-driven change in vibrational level, by separately optimizing the lattice displacement for Δv = -2 and Δv = -1. Second, we improved the quality of the optical pumping light polarization at the atoms by a factor of 5. For the best result, we apply 50 cooling sequences in each direction. The whole cooling sequence takes ~1s.

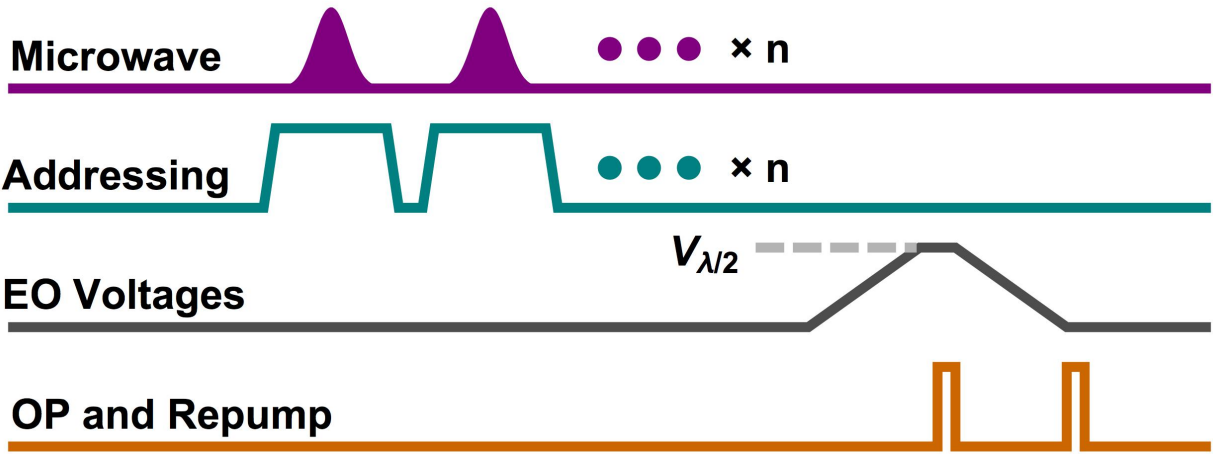

Extended Data Figure 1|Motion Step| A motion step to move n atoms. n atoms are sequentially targeted by the addressing beams and transferred from the stationary state to the motion state using microwaves. The EO voltages are ramped up to move atoms and the motion atoms are optically pumped back to the stationary state. The EO voltages are then ramped back down. A final optical pumping ensures optimal preparation for the next motion step.

**Implementation of a motion step**

Extended Data Figure 1 illustrates our timing sequence for one motion step. Before any motion, atoms are optically pumped to the $|F=4, m_F=-4\rangle$ state (not shown). The atoms to be moved are transferred to the $|F=3, m_F=-3\rangle$ state sequentially. The addressing lasers, directed by micro mechanical electronic systems (MEMS) mirrors, cross at a target atom in the 3D array, AC Stark shifting its resonance frequency between $|F=4, m_F=-4\rangle$ and $|F=3, m_F=-3\rangle$ by -50 kHz with respect to the atoms not in path of either addressing laser. The addressing laser powers are ramped up over 40 us, after which time we wait another 110 us for our intensity lock to settle. We drive the transition in the target atoms with a 3 ms long adiabatic fast passage (AFP) microwave pulse, which involves a 12 kHz frequency sweep. The crosstalk is less than $< 3 \times 10^{-3}$.

To initiate motion, the polarization of one of the lattice beams is linearly rotated by $\pi$ over 3 ms by ramping the voltages on the Pockels cells. The atoms in $|F=3, m_F=-3\rangle$ are then optically pumped (intensity is 40 mW/cm$^2$ and the detuning is 7.5 MHz) to $|F=4, m_F=-4\rangle$ in 0.25ms. The voltages are then ramped back to zero. A final optical pumping step over 0.25 ms ensures that all atoms are back to $|F=4, m_F=-4\rangle$ for the next motion step.

**Measuring state flip fidelity**

To measure the efficiency of our addressing scheme, we take an occupancy map, projection sideband cool the atoms and optically pump them to the $|F=4, m_F=-4\rangle$ state. We then sequentially flip the state of all the atoms within the 5x5x5 region to $|F=3, m_F=-3\rangle$ using targeted addressing. Another laser beam resonant with the $|F=4\rangle$ to $|F'=5\rangle$ transition then pushes away the atoms that were left in the $|F=4, m_F=-4\rangle$ state. A new occupancy map is then taken to identify the atoms that were successfully transferred to $|F=3, m_F=-3\rangle$. Averaging over 50 implementations, we measure a state flip fidelity of 0.986(5). However, the addressing laser beam drifts slowly once aligned, which can decrease the state flip fidelity by about 0.02 after about 100 sorting implementations.

**Measuring motion fidelities**

Motion errors can occur when atoms spontaneously emit lattice light. An atom will usually be lost during motion if light scattering leaves it in the anti-trapped state. Occasionally the atom will site-hop, if it stays trapped but follows the "wrong" lattice potential. We experimentally measure the motion fidelities for atoms in $|F=4, m_F=-4\rangle$ and $|F=3, m_F=-3\rangle$ separately (see Extended Data Figure 2). Atoms are first projection sideband cooled and optically pumped to $|F=4, m_F=-4\rangle$. To find the cumulative effect of making 2$N$ motion steps in a given direction, we ramp up to the Pockels cells' half wave voltage, $V_0$, then ramp down, to -$V_0$, repeating the process $N$ times. No optical pumping or state flips are applied during the motions, so all the atoms move back and forth by one lattice spacing around their initial positions. By comparing the occupancy maps before and after these motions steps, we can identify the percentage of atoms that successfully return to their initial positions, which we call the motion fidelity. For motion in $|F=3, m_F=-3\rangle$, the sequence is the same except that after the atoms are optically pumped to $|F=4, m_F=-4\rangle$, a global microwave pulse is applied to flip the state of all atoms to $|F=3, m_F=-3\rangle$ before executing the motions. Each data point in Extended Data Figure 2 is averaged over 10 implementations and corrected for the loss due to collisions with background gas atoms. A linear fit gives the fidelities per motion step. For atoms in the $|F=4, m_F=-4\rangle$ state and motion in the $\{x, y, z\}$ lattice directions, the fidelities are: {0.9951(6), 0.9982(6), 0.9962(4)}. The corresponding fidelities for atoms in the $|F=3, m_F=-3\rangle$ state are: {0.9956(4), 0.9961(10), 0.9956(1)}. The calculated probability of spontaneous emission for an atom in the vibrational ground state during a motion step is $3.5\times10^{-3}$.

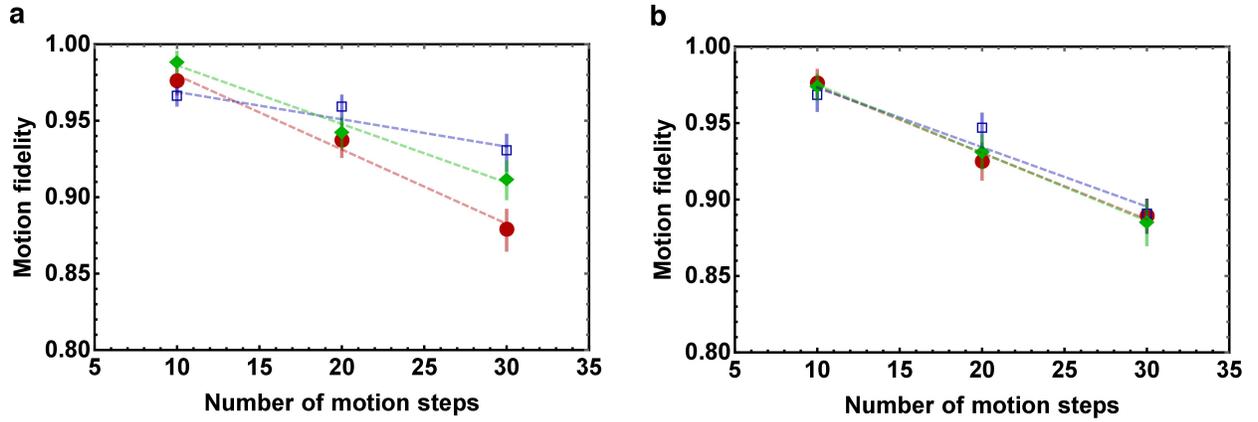

**Extended Data Figure 2|Motion Fidelities|** Measured motion fidelity as a function of number of motion steps in a, |F = 4, mF = -4⟩ and b, |F = 3, mF = -3⟩ for x (maroon circles), y (blue squares) and z (green diamonds). The lines are fits to the data.

**Sorting Algorithm**

We have generalized the sorting algorithm for any initial $N \times N \times N$ lattice and any final $i \times j \times k$ sub-lattice. If $i = j = N$, then only balancing and compaction steps are needed. If $i, j < N$, then extra motion steps in $x$ and $y$ are added to move as many atoms as possible in to $i \times j \times N$ sub-lattice from "outside" before balancing and compaction, where outside is the full lattice minus the target sub-lattice. For example, to fill a 4x4x3 sub-lattice, as many atoms as possible are first moved from into a 4x4x5 region in two motion steps, one in each x and y, from outer y-z and x-z planes of the lattice. Balancing and compaction are then applied to a 4x4x5 lattice rather than a 5x5x5 lattice. The simulations we describe below suggest that, even though this procedure does not always empty the outside planes, there are always enough atoms to fill a 4x4x3 sub-lattice starting from a 50% filled 5x5x5 lattice.

The steps for balancing a $i \times j \times N$ lattice to fill an $i \times j \times k$ sub-lattice are roughly as follows:

1. If this is the first iteration, choose a dividing plane, P, to be an x-z plane. Otherwise, choose a dividing plane, P, to be perpendicular (either x-z or y-z) to the previous iteration. Choose P to divide the lattice into two parts, $S_1$ and $S_2$, that are as equal as possible (i.e., a difference of one plane is permitted if the lattice dimension is odd).
2. If the number of z-rows in $S_1$ ($S_2$) is $n$ ($m$), the required number of atoms in $S_1$ ($S_2$) is $k*n$ ($k*m$). Move atoms between the two sub-lattices until they each have at least the required number.
3. Repeat these steps for $S_1$ and $S_2$ separately, stopping when each is just a single z-row.

Balancing guarantees that there are $k$ atoms in each of the $i*j$ z-rows. These atoms are then moved in the $z$ direction ("compaction") in parallel in order to fill the desired $k$ planes, usually in the middle of the accessible lattice. The algorithm minimizes the number of motion steps.

The sorting algorithm can probably be improved by replacing the initial steps to empty the outer *x-z* and *y-z* planes by a more optimal algorithm. For instance, the first sort could be modified to distribute the extra atoms evenly and thus reduce the number of correction steps.

**Monte-Carlo simulations**

Monte-Carlo simulations of this sorting algorithm start with a randomly half-filled 3D array. Errors are probabilistically applied at each motion step and atom loss is applied after completion of a sort. There is a separate motion fidelity for each internal state, which we take to be the average of the measured fidelities in the three directions. One thousand simulations were run for various lattice dimensions and various target sub-lattices. For filling a 5x5x2 or a 4x4x3 sub-lattice from a half-filled 5x5x5 lattice, the simulations predict an average filling factor of ~ 0.97 after three sorts, in agreement with our measured filling factor to within the uncertainty associated with our measured errors.

**Real-time Control**

The implementation of sorting requires changing the timing sequence on the fly. This is accomplished by combining real-time data analysis with two Field Programmable Gate Arrays (FPGAs). The experiment has a "back-bone" of a fixed timing sequence. After the motion-steps have been generated based on the initial occupancy map, the FPGAs pause that fixed timing sequence and take control of the electronic channels (optical pumping, Pockels cell voltages, addressing, microwaves) required for sorting. The data for sorting, which is a sequence of directions for the motion steps as well as the lattice sites to be addressed for each motion step, is communicated to the FPGAs by the program that generates the occupancy map and creates the sorting plan. The FPGAs convert the motion steps into several voltage sequences that are output synchronously. After the motion steps have been executed, the FPGAs transfer the timing control back to the fixed "back-bone", which resumes where it was paused.